# Long-range quantum entanglement in dielectric mu-near-zero metamaterials

Olivia Mello[1], Larissa Vertchenko[2], Seth Nelson[3], Adrien Debacq[4], Durdu Guney[1,5], Eric Mazur[1], Michaël Lobet[1,4,*]

[1]John A. Paulson School of Engineering and Applied Sciences, Harvard University, 9 Oxford Street, Cambridge, MA 02138, USA
[2] Sparrow Quantum, 2100 Copenhagen, Denmark
[3]Physics Department, Michigan Technological University, Houghton, Michigan, USA
[4] Department of Physics and Namur Institute of Structured Materials, University of Namur, Rue de Bruxelles 51, 5000 Namur, Belgium
[5] Electrical & Computer Engineering Department, Michigan Technological University, Houghton, Michigan, USA

*Corresponding author: michael.lobet@unamur.be

*Emails of the authors:*

Olivia Mello: oliviamello@g.harvard.edu
Larissa Vertchenko: Larissa.Vertchenko@sparrowquantum.com
Seth Nelson: srnelson@mtu.edu
Adrien Debacq: adrien.debacq@unamur.be
Durdu Guney: dguney@mtu.edu
Eric Mazur : mazur@seas.harvard.edu
Michaël Lobet: michael.lobet@unamur.be

## Abstract

Entanglement is paramount in quantum information processing. Many quantum systems suffer from spatial decoherence in distances over a wavelength and cannot be sustained over short time periods due to dissipation. However, long range solutions are required for the development of quantum information processing on chip. Photonic reservoirs mediating the interactions between qubits and their environment are suggested. Recent research takes advantage of extended wavelength inside near-zero refractive index media to solve the long-range problem along with less sensitivity on the position of quantum emitters. However, those recent proposals use plasmonic epsilon near-zero waveguides that are intrinsically lossy. Here, we propose a fully dielectric platform, compatible with the Nitrogen Vacancy (NV) diamond centers on-chip technology, to drastically improve the range of entanglement over 17 free-space wavelengths, or approximatively 12.5μm, using mu near-zero metamaterials. We evaluate transient and steady state concurrence demonstrating an order of magnitude enhancement compared to previous works. This is, to the best of our knowledge, the first time that such a long distance is reported using this strategy. Moreover, value of the zero time delay second order correlation function $g_{12}^{(2)}(0)$ are provided, showing antibunching signature correlated with a high degree of concurrence.

# 1 Introduction

A principal topic in quantum information processes is generating and maintaining entanglement[1]. Entanglement, which is defined as the non-separability of quantum states, is important for quantum teleportation, quantum metrology, quantum communication, and quantum computing. First demonstrated in optical and atomic/ionic systems[1,2], entanglement has now been demonstrated in a variety of solid state systems including for Josephson junctions[3], the spin of quantum dots[4], and in single photon emission from silicon carbide[5]. However, in solid-state on-chip systems, entanglement is observed over distances within only one spatial wavelength or not much greater[6]. In order to have information transfer over long distances, entanglement and interactions between qubits must be maintained over large spatial separations avoiding decoherence from coupling of the quantum emitters and its environment[7].

It has been shown that while typically spontaneous emission is a form of decoherence in most entanglement regimes, because it is a source of dissipation into the environment, this dissipation can also be a source of entanglement generation[8,9] for two level systems interacting in a bosonic field. This is particularly relevant for the case of cooperative enhancement of spontaneous emission via Dicke states[10]. Dicke states correspond the symmetric and antisymmetric linear combinations of the ground $|g\rangle$ and excited states $|e\rangle$ for a collection of two-level systems. For a set of two qubits, the symmetric superradiant $|+\rangle = (|e\rangle|g\rangle + |g\rangle|e\rangle)/\sqrt{2}$ and antisymmetric subradiant $|-\rangle = (|e\rangle|g\rangle - |g\rangle|e\rangle)/\sqrt{2}$ Dicke states are the maximally entangled states in the system as we cannot factor them. As cooperative spontaneous emission mediates interactions with these states, it is one means of entanglement generation for such systems.

There have been multiple proposals for long-distance entanglement, such as the Duan-Lukin-Cirac-Zoller (DLCZ) protocol between two distant cold atom cavity ensembles[11]. Long distance heralded entanglement with optical photons has been experimentally demonstrated with distant quantum dot hole-spins[12] and with electron spins in diamond nitrogen vacancy centers[13]. Such protocols are difficult to attain on an on-chip system because they require, for example, complicated entanglement preparation protocols involving entanglement swapping[14] and a separate intermediary location between the two qubits. The cooperative quantum optical effects that generate entanglement on a mesoscopic (on-chip) scale rapidly depreciate after length scales greater than one interaction wavelength ($kr < 1$ for having cooperative effects, with $k$ the wavenumber, $r$ the position and $\lambda$ the wavelength inside the material). There have been several proposals to use photonic reservoirs to mediate the interactions between qubits as well as their environment[15].

In that context, the development of near-zero refractive index (NZI) photonics is of prime interest[16–18]. Indeed, inside such NZI media, the effective wavelength $\lambda = \frac{\lambda_0}{n}$ stretches to infinity while the wavevector goes to zero[19], therefore quantum emitters are able to cooperate because of enhanced coherence length. As the refractive index is defined by $n = \pm\sqrt{\varepsilon\mu}$, the NZI state can be reached by 3 different ways: either the electric permittivity $\varepsilon$ is close to zero – the epsilon near-zero (ENZ) category –, the magnetic permeability $\mu$ is close to zero – the mu near-zero (MNZ) category – or both $\varepsilon$ and $\mu$ are close to zero – the epsilon and mu near-zero (EMNZ) category[20].

Quantum entanglement mediated by ENZ reservoirs[8,21–24] have been recently discussed as systems conferring both local field enhancement and constructive interference due to the relaxed phase matching conditions[25] at the ENZ frequency. Those are plasmonic waveguide systems which operate at the cutoff frequency to generate the ENZ behavior. Despite great interest, those plasmonic systems are intrinsically lossy. To tackle this problem, based on all-dielectric NZI metamaterials[18,26–28], we previously designed diamond ENZ metamaterials platform that theoretically and numerically showed superradiant decay rate enhancement over distances greater than 13 times the free-space wavelength[29]. A power enhancement of three orders of magnitude higher than an incoherent array of emitters in bulk diamond was observed, corresponding to an $N^2$ scaling with the number of emitters as a characteristic of superradiance. While interesting from a low-loss point of view, this 2D ENZ platform presents a natively high impedance $Z = \sqrt{\frac{\mu}{\varepsilon}}$ as we approach the NZI frequency, hence hindering in and out-coupling with the structure. Nevertheless, as recently shown, both the category of NZI and the spatial dimensions of the photonic media have a drastic impact on fundamental radiative processes such as spontaneous emission[30]. Therefore, investigating other NZI categories could potentially solve this impedance problem.

Here, we demonstrate how planar MNZ metamaterials can significantly improve the range of entanglement over more than 10 free-space wavelengths. This MNZ design has a transverse electric (TE) polarization with a finite impedance. For systems of entangled qubits in emitter-based systems on-chip, our system represents a significant improvement in entanglement over distance between the emitters. Additionally, the method of entanglement generation in this system is comparatively quite simple, based on the cooperative spontaneous emission between emitters that we appreciate due to the significantly reduced phase advance in near-zero refractive index materials.

In our theoretical model, we utilize the calculated [29]coupling coefficients and decay rates in the Lindblad master equation to describe the quantum dynamics of a two-qubit system through its density matrix. We provide the general coupled differential equations for the density matrix elements of the system in the presence of pumps to describe the full quantum dynamics of the two-qubit system under different initial conditions. These equations can be readily used to implement various quantum tasks such as entanglement generation, quantum logic gates, quantum teleportation, and other two-qubit quantum operations in metamaterials and photonic crystals akin to [31,32].

# 2 Theoretical framework of entanglement in two-level systems

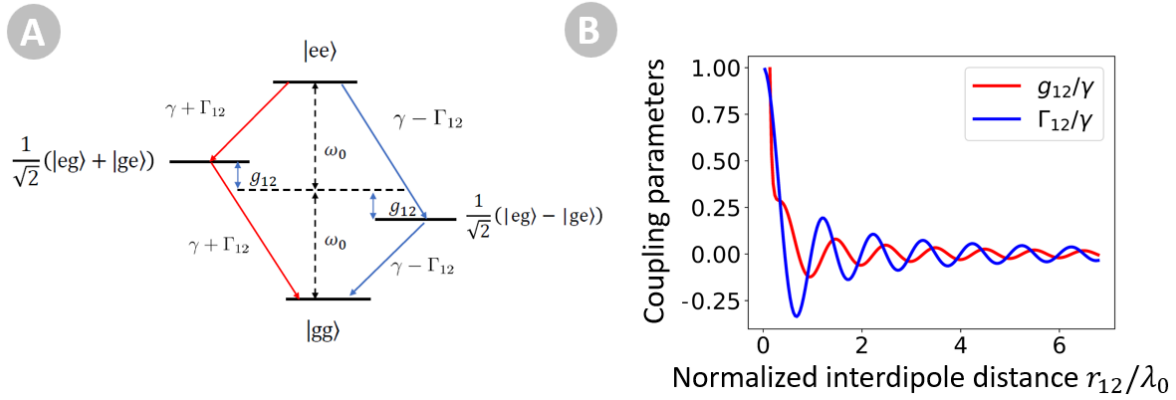

**Figure 1.** (a) System of two two-level atoms with symmetric and antisymmetric states $|+\rangle$ and $|-\rangle$ with cooperative decay rates $\gamma \pm \Gamma_{12}$ and cooperative frequency level shifts $g_{12}$. (b) Plots of the normalized cooperative decay rate $\Gamma_{12}$ and the dipole-dipole couplings $g_{12}$ in free space, as a function of the normalized interdipole distance.

The two-level atomic system considered here is represented at Figure 1a with the corresponding Hamiltonian describing the coherent part of the dynamics between two emitters (Fig. 1a) being

$$H = \sum_i \hbar(\omega_0 + g_{ii})\sigma_i^\dagger \sigma_i + \sum_{i \neq j} \hbar g_{ij}\sigma_i^\dagger \sigma_j \tag{1}$$

where $g_{ii}$ represents the photonic Lamb shift[22,33] generated as a result of self-interaction with a single emitter. The Lamb shift arises when the bound electrons interact with fluctuations in surrounding vacuum field in the medium. This shift is typically incorporated into the transition frequency term of any chosen emitter and thus does not affect our calculations here. $\sigma_{i,j}$ are the ladder operators between excited states $|e_{1,2}\rangle$ and ground states $|g_{1,2}\rangle$ for emitters 1 and 2. The coherent dipole-dipole coupling term $g_{ij}$ represents interactions that result in a frequency shift between emitters. Fig 1.a illustrates how the dipole-dipole coupling term affects the frequency shift in the symmetric and antisymmetric states in the system of two two-level atoms. In resonant structures we can calculate this term as the real part of the Green's tensor $\overleftrightarrow{G}(\boldsymbol{r}_i, \boldsymbol{r}_j, \omega_j)$ [22,29,34],

$$g_{ij} = \frac{2\omega_j^2}{\hbar \varepsilon_0 c^2} \boldsymbol{d}_i Re[\overleftrightarrow{G}(\boldsymbol{r}_i, \boldsymbol{r}_j, \omega_j)]\boldsymbol{d}_j^* \tag{2}$$

where $\boldsymbol{d}_{i,j}$ are the dipole moments of the *i*th and *j*th emitters at positions $\boldsymbol{r}_i$ and $\boldsymbol{r}_j$ for a frequency $\omega_j$.

The collective spontaneous emission rates $\gamma_{ij}$, defined as $\Gamma_{12}$ for the two-atom system (Fig 1.a), arise from coupling between the atoms through the vacuum field. The cooperative decay rates $\gamma_{ij} = \Gamma_{12}$ and the single atom spontaneous decay rate $\gamma_{ii} = \gamma$ are determined by the imaginary component of the Green's function:

$$\gamma_{ij} = \frac{2\omega_j^2}{\hbar\varepsilon_0 c^2} d_i Im[\overleftrightarrow{G}(r_i, r_j, \omega_j)] d_j^* \quad (3)$$

It should be noted that the cooperative decay rate $\gamma_{12} = \Gamma_{12}$ and single atom spontaneous decay rate $\gamma_{ii}$ factor in when calculating the density matrix as they are the dissipative components in the Lindblad master equation[34,35]:

$$\frac{\partial \rho}{\partial t} = \frac{1}{i\hbar}[H,\rho] - \frac{1}{2}\sum_{i,j}^{2} \gamma_{i,j}\left(\rho\sigma_i^\dagger \sigma_j + \sigma_i^\dagger \sigma_i \rho - 2\sigma_i \rho \sigma_j^\dagger\right) \quad (4)$$

with $\sigma_i = |g_i\rangle\langle e_i|$ is the ladder operators for the $i$th emitter.

In free-space or inside a homogeneous medium with a wavevector $k$, the cooperative decay rate and the dipole-dipole coupling can be evaluated analytically[36,37] as

$$\gamma_{i,j} = \frac{3\gamma}{4}([1-(\bar{d}\cdot\bar{r}_{ij})^2])\frac{\sin(kr)}{kr} + ([1-3(\bar{d}\cdot\bar{r}_{ij})^2])\left[\frac{\cos(kr)}{(kr)^2} - \frac{\sin(kr)}{(kr)^3}\right] \quad (5)$$

$$g_{i,j} = \frac{3\gamma}{4}(-[1-(\bar{d}\cdot\bar{r}_{ij})^2])\frac{\cos(kr)}{kr} + ([1-3(\bar{d}\cdot\bar{r}_{ij})^2])\left[\frac{\sin(kr)}{(kr)^2} + \frac{\cos(kr)}{(kr)^3}\right] \quad (6)$$

where $\gamma$ is the free space spontaneous emission rate and $\bar{d}$ and $\bar{r}$ are unit vectors along the atomic transition dipole moments and the vector $r_{ij} = r_j - r_i$, respectively.

Fig 1.b shows the plots for the cooperative dissipative coupling for two dipoles $\Gamma_{12}$ and the coherent dipole-dipole coupling term $g_{12}$ over a normalized interdipole separation in free space. As the separation between the two dipoles exceeds one wavelength for the free space case, both the cooperative decay and dipole-dipole coupling terms rapidly degrade. We will observe how these two terms affect entanglement and how we can engineer them to sustain entanglement over extents greater than 10 free-space wavelengths or more.

Introduced by Wooters[38], concurrence helps to characterize the extent to which two emitters are entangled with one another. The concurrence $C$ is defined as

$$C = max(0, \sqrt{\lambda_1} - \sqrt{\lambda_2} - \sqrt{\lambda_3} - \sqrt{\lambda_4}) \quad (7)$$

where $\lambda_i$ are the eigenvalues of the matrix $\rho\tilde{\rho}$ in the descending order and $\tilde{\rho} = (\sigma_y \otimes \sigma_y)\rho^*(\sigma_y \otimes \sigma_y)$ and $\sigma_y$ is the 2-by-2 Pauli matrix. Therefore, for completely unentangled states, $C = 0$ and for a maximally entangled system, $C = 1$. To calculate the concurrence, we must solve the time-dependent density matrix components $\rho(t)$ with the Lindblad master equation (equation 4). We solve it in the Dicke basis: $|0\rangle = |g_1\rangle|g_2\rangle, |3\rangle = |e1\rangle|e2\rangle, |+\rangle = (|e_1\rangle|g_2\rangle + |g_1\rangle|e_2\rangle)/\sqrt{2}$ and $|-\rangle = (|e_1\rangle|g_2\rangle - |g_1\rangle|e_2\rangle)/\sqrt{2}$.

Up until this point we have considered the emitter decay and dipole-dipole coupling due to its interaction with the electromagnetic field of the photonic reservoir. However, depending on the emitter, there can be other non-radiative channels for decay, known as dephasing. The master equation can additionally concern itself with dephasing $\gamma'$ [8,35,39]. The contribution of dephasing to the dynamics of the master equation is

$$L\rho = \sum_{i=1,2} \frac{\gamma'}{2}[2\sigma_i^\dagger \sigma_i \rho \sigma_i \sigma_i^\dagger - \sigma_i \sigma_i^\dagger \sigma_i^\dagger \sigma_i \rho - \rho \sigma_i \sigma_i^\dagger \sigma_i^\dagger \sigma_i] \tag{8}$$

For example, in the silicon vacancy in diamond, several sources are responsible of dephasing. A strain-induced increase of orbital splitting to an energy at which phonon population is one contribution[40]. Phonon excitations in the diamond lattice from the lower to the upper orbital branches in the ground-state manifold is the largest contribution. Dephasing in the silicon vacancy center is highly dependent on temperature, making dilution refrigeration[40] an often important aspect to measurement. For the sake of our calculations, we assume a system at low temperatures such that the dephasing terms typically present in the master equation are small. With these considerations, we can calculate the terms $\partial \rho_{ij}/\partial t$ in the Dicke basis.

To solve this set of equations, we assume a single initial excitation where we prepare the initial unentangled state, $|e_1, g_2\rangle = \frac{1}{\sqrt{2}}(|+\rangle + |-\rangle)$. This gives us the initial conditions for the density matrix at $t = 0, \rho_{++}(0) = \rho_{--}(0) = \rho_{+-}(0) = \rho_{-+}(0) = 1/2$, and $\rho_{33}(0) = \rho_{00}(0) = 0$. This greatly simplifies our set of differential equations, and the time-dependent solution for the nonzero elements of the density matrix become

$$\begin{cases} \rho_{++}(t) = \frac{1}{2}e^{-(\gamma+\Gamma_{12})t} \\ \rho_{--}(t) = \frac{1}{2}e^{-(\gamma-\Gamma_{12})t} \\ \rho_{+-}(t) = \frac{1}{2}e^{-(\gamma-2ig_{12})t} \\ \rho_{-+}(t) = \frac{1}{2}e^{-(\gamma+2ig_{12})t} \end{cases} \tag{9}$$

If the density matrix of this system only contains the four above elements, the expression of the concurrence can be significantly simplified[8] to

$$C(t) = \sqrt{(\rho_{++} - \rho_{--})^2 + 4Im(\rho_{+-})^2}. \tag{10}$$

With the density matrix solutions above, the expression for the concurrence reduces to

$$C = \frac{1}{2}\sqrt{(e^{-(\gamma+\Gamma_{12})t} - e^{-(\gamma-\Gamma_{12})t})^2 + 4e^{-2\gamma t}\sin^2(2g_{12}t)}. \tag{11}$$

For a configuration of two emitters in free space, we observe a transient concurrence that rapidly degrades as soon as the two emitters are separated as little as half a wavelength. We can observe from the expression of the concurrence that for ideal concurrence values, $\Gamma_{12}/\gamma = 1$ for large inter-emitter separations and $\Gamma_{12} \gg g_{12}$. As we demonstrated in previous work[29], near-zero index metamaterials, particularly ENZ metamaterials, exhibit a strong spatial enhancement in cooperative decay rate. Here, we demonstrate that MNZ metamaterials display cooperative enhancement in the decay terms $\Gamma_{ij}$ over large spatial extents. This is because, while the retrieved effective permittivity $\varepsilon_r$ is a nonzero value, the effective index $n_\text{eff}$ still approaches the near-zero limit at the wavelength in which $\mu_r$ crosses zero.

## 3 Design of the MNZ metamaterial

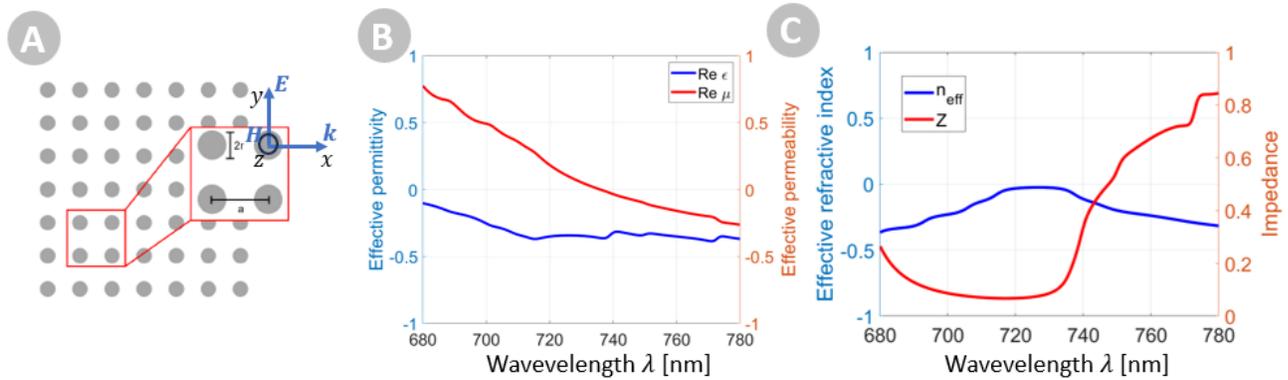

**Figure 2.** (a) Schematic of the 2D square metamaterial lattice with a pitch $a$ and radius $r$. **(b)** Effective parameters $\varepsilon_r$ and $\mu_r$ with a $\mu$ zero crossing at 737 nm. **(c)** The retrieved effective refractive index $n_{eff}$ and impedance $Z$.

Recently, the development of NZI photonics showed a range of particularly interesting effects for light and thermal emission, nonlinear optics, sensing applications, and time-varying photonics[20]. This includes, for example, applications such as supercoupling[16], cloaking[19,41,42], crosstalk prohibition[43], photonic doping[44], or impurity-immunity[45].

We start by numerically designing a planar MNZ metamaterials composed of square lattice structure of period $a = 505\ nm$ with a pillar radius $r = 115\ nm$ (Fig 2a). The quantum emitter will be based on silicon vacancies (SiVs) in diamond, with a corresponding zero-phonon line at 737 nm. Excitation comes from a fundamental transverse electric (TE) mode source. The MNZ mode is TE polarized, with the field $H_z$ polarized out of plane.

Although the period $a$ in this case is not much smaller than the operating wavelength $\lambda_0 = 737\ nm$ in vacuum, it should be noted that the effective wavelength $\lambda = \lambda_0/n_\text{eff}$ inside the MNZ metamaterial is bigger than $a$ once the effective refractive index is close to zero. Therefore, we follow the formalism of the homogenization criteria in the limit of short optical wavelengths[46]. The effective refractive index $n_\text{eff}$ can be retrieved by recording the average change in phase between each pillar in the simulation region while the impedance $Z$ and the effective constitutive parameters $\varepsilon_r$ and $\mu_r$ are retrieved by calculating the transfer matrix[29,47–50] using the transmitted and reflected fields recorded by the monitors near the boundaries of the metamaterial.

Figure 2(b) shows that, as imposed by design, the real part of the effective permeability crosses zero at 737 nm, the zero-phonon line of the silicon vacancy center, while the effective permittivity $\varepsilon_r = -0.36$. Figure 2(c) shows the corresponding minimum absolute value of the effective phase index $n_{\text{eff}} = -0.03$ at the wavelength of the $\mu$ zero crossing. This corresponds to a low value of impedance[27], as $Z = \sqrt{\frac{\mu}{\varepsilon}} = 0.2$. The reason that the value of $n_{\text{eff}}$ never fully crosses zero is that the retrieved optical parameters $\varepsilon_r$ and $\mu_r$ have small imaginary components (see SI for $Im(\varepsilon_r)$, $Im(\mu_r)$ and the impact on $Im(n_{\text{eff}})$). This further indicates low losses from these imaginary contributions.

Furthermore, to double-check the consistency of the retrieved parameters using the parameter retrieval methods, we calculated the effective index calculated by averaging the phase advance between each pillar for 10 periods of the metamaterial using Ansys Lumerical FDTD. This agreement (see Fig S3 in SI) confirms the validity of the photonic crystal as a MNZ metamaterial structure.

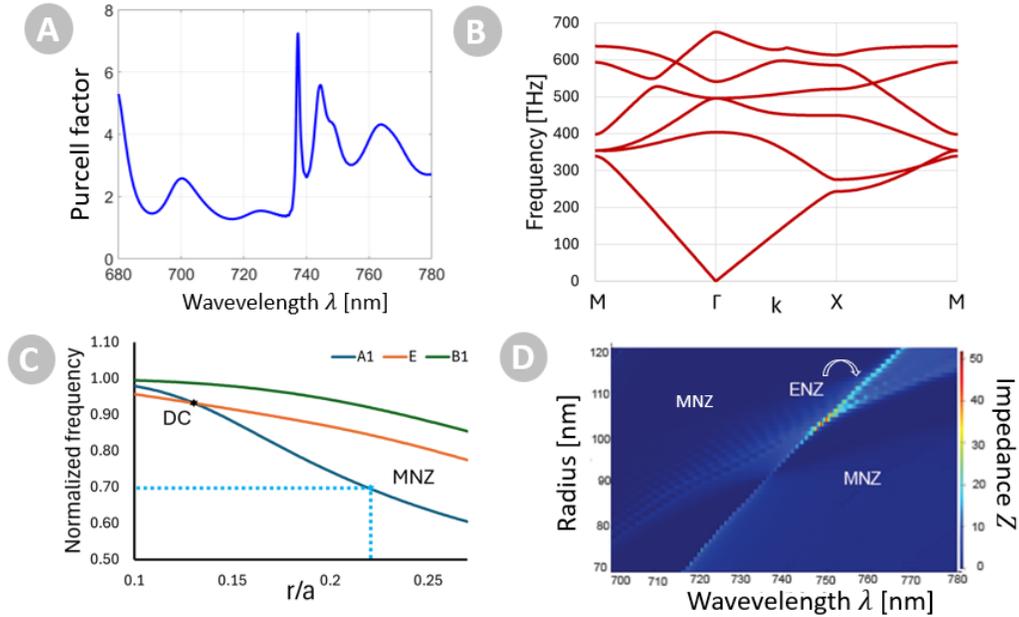

**Figure 3. (a)** Spectrum of the Purcell factor in the MNZ structure, with a peak near 737 nm. **(b)** Band structure of the 2D square lattice of dielectrics pillars ($n = 2.4064$) with a pitch $a = 505\ nm$ and radius $r = 115\ nm$. **(c)** Plot of the eigenfrequency at the $\Gamma$-point of the A1 (monopolar) mode, the two degenerate E (dipolar) modes, and the B1 (quadrupolar) mode as we change the radius from the EMNZ design value (Dirac cone). **(d)** Plot of the impedance spectrum with various radius showing the MNZ (blue) and ENZ (bright diagonal) regions.

The low impedance justifies the choice of a MNZ metamaterial compared to an ENZ metamaterial[29]. The ENZ designs[19,30], while demonstrating a high Purcell enhancement relative to the large mode volumes that near-zero index modes occupy, cannot couple effectively in or out of plane. This lack of efficient coupling is experimentally detrimental. In contrast, the MNZ design gives a lower but non-zero impedance, making it much easier to couple radiation in or out of the material both in or out of plane (see discussion in SI for quantification and distinction with an EMNZ case). This coupling possibility is a particular advantage when working with low-index materials such as diamond, as they require either suspended or angle-etched waveguides to efficiently couple out of the metamaterial without radiation leaking into the substrate. A modest Purcell enhancement at the $\mu$ near-zero crossing at 737 nm is still present ($F_P = 7$) as indicated in Figure 3a, because we are still operating at a shallow parabolic band edge at the center of the Brillouin zone Γ (Figure 3b).

The MNZ structure is derived from an EMNZ structure with a triply-degenerate Dirac cone dispersion[17,18,26,41,51] coming from the degeneracy of the monopolar $A1$ mode and the doubly-degenerated dipolar $E$ modes[51]. The original 2D EMNZ diamond metamaterial structure has a pitch $a = 684\ nm$ and radius $r = 87\ nm$. Figure 3c shows how detuning the radius and re-scaling the pitch, we can break the degeneracy of these three bands (Dirac cone occurring at the normalized frequency 0.92 for $r/a = 0.13$) and again operate near a band edge after breaking the degeneracy of these modes (MNZ mode occuring at the normalized frequency 0.7 for $r/a = 0.22$) . In Figure 3d, we similarly see how breaking the degeneracy of the modes by either increasing or decreasing the radius controls the impedance, as we are changing the values of $\varepsilon$ and $\mu$ relative to one another. Near the spike of impedance for shrunken pillars, we see a theoretical ENZ behavior (bright dots in Fig. 3d). However, in our simulations we do not see any local or cooperative enhancement at this point because of the existence of an additional band, the flat band, at this point. As described in detail in[41,51], the flat band has propagation that is longitudinal at that point, which causes destructive interference in the fields. It is specifically due to the TE polarization with the $H_z$ monopole mode that we observe MNZ behavior instead of ENZ as is the case for the TM-polarized ENZ structure in Ref. [29].

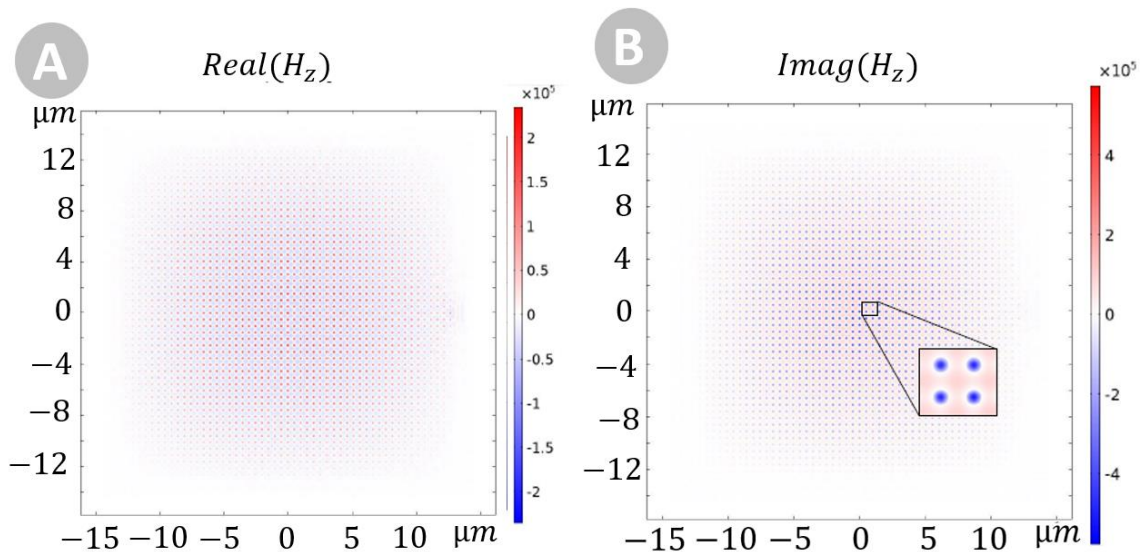

**Figure 4.** (a) The real and (b) imaginary components of the $H_z$ field for the MNZ monopole mode.

Figure 4 shows the real and imaginary components of the MNZ TE modes for a $51 \times 51$ pitch design extending over approximately 12 μm. Both the real and imaginary components of the field extend nearly throughout the metamaterial, demonstrating "supermode"-like behavior where the metamaterial generates larger-scale fundamental Fabry-Pérot cavity mode[22,29,52].

Figure 4(b) demonstrates that the MNZ mode is the monopolar $A1$ mode with a TE-polarization that corresponds to the magnetic resonance that generates the $\mu$ near-zero crossing. The imaginary component of the $H_z$ field, which is proportional to $\Gamma_{12}$, is roughly three times larger in magnitude than the real part of $H_z$, which is proportional to the dipole-dipole coupling $g_{12}$. Having a $\Gamma_{12} \gg g_{12}$ is ideal for attaining a large transient concurrence as determined by the Hamiltonian of our system. We further investigate the long-range entanglement properties of the designed MNZ metamarials here after.

## 4 Entanglement in the MNZ MM

Having the MNZ metamaterial designed allows us to calculate dipole-dipole coupling $g_{12}$ (eq. 2) and the cooperative decay rate $\gamma_{12}$ (eq. 3). We use full-wave numerical simulations (COMSOL Multiphysics) by sweeping a magnetic dipole source from the origin to 25 pitches away along the $x$-direction. As the dipole moves from pillar to pillar along the $X - M$ direction, the real and imaginary components of the $H_z$ field from the TE monopole mode are recorded. As observed in Figure 4, the fields are radially isotropic upwards of 8 µm, therefore transferring that isotropic property to both cooperative enhancement and dipole-dipole coupling.

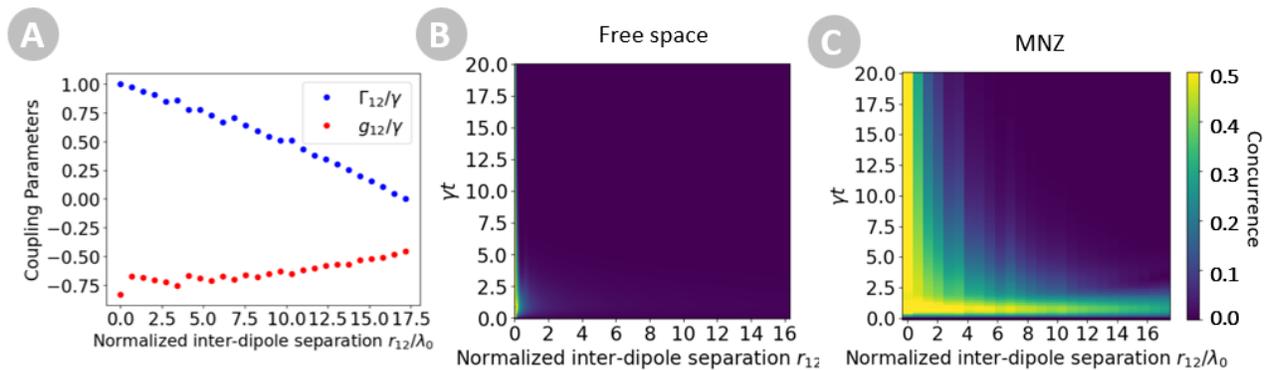

**Figure 5. (a)** Normalized cooperative decay coupling $\Gamma_{12}$ and dipole-dipole coupling $g_{12}$ calculated from moving the dipole source across MNZ structure. **(b)** Plot of the transient concurrence as a function of the normalized interdipole separation for free space and **(c)** the MNZ metamaterial.

Figure 5a demonstrates the cooperative enhancement $\Gamma_{12}/\gamma$ and dipole-dipole coupling $g_{12}/\gamma$, both normalized by the spontaneous decay rate (calculated from pillar to pillar) in the MNZ metamaterial design. For the MNZ material, we use magnetic dipole sources with the dipole moment aligned along the length of the pillars. Using such a source in simulations corresponds to the TE polarization of the MNZ material with the $H_z$ mode aligned along the axis of the pillars. By comparing with Figure 1b (free space), one can observe extended cooperative enhancement for over an order of magnitude longer inter-emitter separations inside the MNZ metamaterial. This enhanced cooperative enhancement $\Gamma_{12}/\gamma > 0$ is sustained for over 17 free-space wavelengths ($\lambda_0 = 737nm$), so about 12.5 µm, which is one order of magnitude higher than in the rolled-up or rectangular ENZ waveguides ($1.5r/\lambda_0$) [23]. Moreover, the coherent dipole-dipole coupling terms $g_{12}/\gamma$ are smaller than that of the free-space dipole-dipole coupling terms which diverge for short inter-emitter separations. $g_{12}/\gamma$ has the same trend as the ENZ waveguides described in [24] with more negative values over an extended range.

With the optical coupling parameters $\Gamma_{12}$ and $g_{12}$ at hand, we can calculate the transient concurrence from equation 11. Figures 5b and c show the plot of transient concurrence over a normalized time $\gamma t$ and over a wavelength-normalized inter-emitter separation for free space and in the MNZ structure. Note that in both figures the concurrence at $\gamma t = 0$ is $C(0) = 0$ as there has yet to be a spontaneous decay event to induce entanglement via cooperative decay through the symmetric superradiant state.

Shortly after a spontaneous decay event, a high concurrence is maintained throughout the MNZ structure (Fig. 5c), in contrast to free space, where the concurrence goes to zero almost immediately. There is robust entanglement shortly after $\gamma t = 0$ that persists for spatial separations for over 20 times the extent for the case of the concurrence in a vacuum system. This temporal variation is similar to previous work[23]. The concurrence remains high, upwards of 0.35, for even 17 wavelengths, or roughly 12.5 µm. This distance of high concurrence entanglement is nearly an order of magnitude longer than the one in the ENZ plasmonic waveguides[21,23]. Not only is there a drastic spatial improvement of entanglement in the MNZ structure compared to in free space (Figure 5b), but there is also additionally a significant temporal enhancement in the concurrence as it decays with time.

As the concurrence is transient, after a long period of time the system needs to be sustained by an external source to prolong the entanglement. To remedy this, we can apply an external pump source to both qubits to observe steady state behavior. For a pump potential

$$V = -\sum_{i=1}^{2} \hbar\left(\Omega_i e^{-i\Delta_i t}\sigma_i^\dagger + \Omega_i^* e^{i\Delta_i t}\sigma_i\right), \quad (12)$$

we add a term $\frac{1}{i\hbar}[V,\rho]$ to the right-hand side of the master equation in equation 4. The effective Rabi terms are $\Omega_i$ for the *i*-th emitter in the MNZ medium. The parameter $\Delta_i = \omega - \omega_p$ is the pump detuning term for a pump frequency $\omega_p$. For the sake of the following calculations, we assume that the pump operates at the same frequency as the resonant frequency of the qubits. This potential term significantly complicates the equations for the time evolution of the density matrix $\rho(t)$. It is preferable to solve this set of equations in this basis of $|0\rangle = |g1,g2\rangle, |1\rangle = |g1,e2\rangle, |2\rangle = |e1,g2\rangle, and |3\rangle = |e1,e2\rangle$. Using the standard basis and the potential term in equation 4 and assuming a zero detuning terms $\Delta_i = 0$, we obtain a system of 16 equations (See SI).

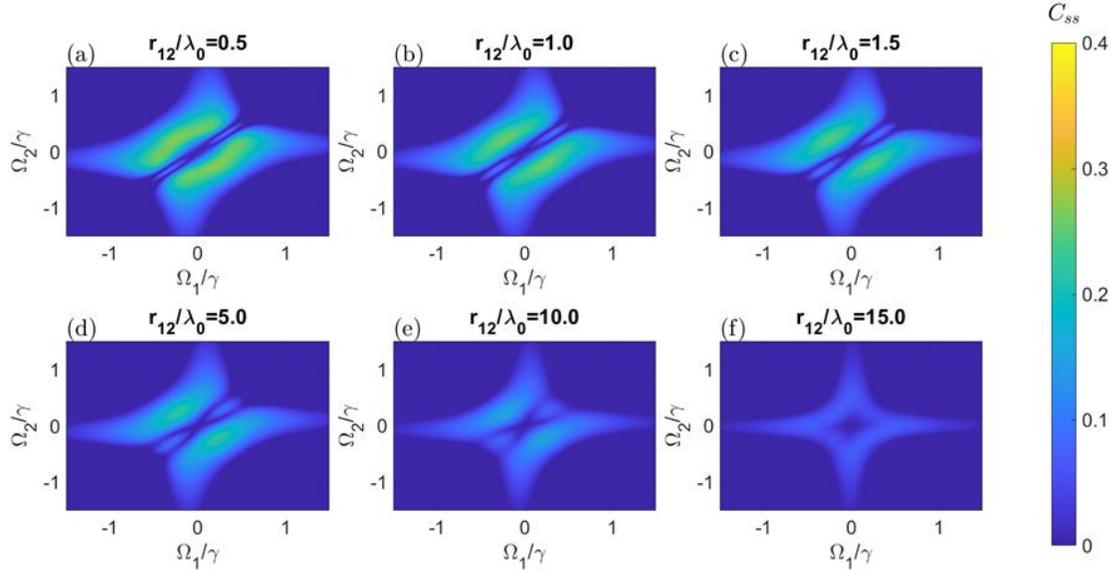

**Figure 6**. Steady state concurrence as a function of normalized Rabi frequencies $\Omega_1/\gamma$ and $\Omega_2/\gamma$ corresponding to two dipoles embedded inside the MNZ metamaterial and separated by (**a**) $0.5\lambda_0$, (**b**) $\lambda_0$, (**c**) $1.5\lambda_0$, (**d**) $5\lambda_0$, (**e**) $10\lambda_0$, and (**f**) $15\lambda_0$, assuming a normalized time of $\gamma t = 90$ for the steady-state[20]. The coupling parameters used are given in Fig. 5a.

Figure 6 shows the steady state concurrence $C_{ss}$ as a function of normalized Rabi frequencies $\Omega_1/\gamma$ and $\Omega_2/\gamma$ for the two quantum emitters embedded inside the MNZ metamaterial and separated by different distances $r_{12}$ ranging from $0.5\lambda_0$ (Fig. 6a) to $15\lambda_0$ (Fig. 6f). The maximum steady-state concurrence $[(C_{ss})]$ for each separation occurs at an antisymmetric pumping configuration of $\Omega_1 = -\Omega_2$, except for $15\lambda_0$ where the $(C_{ss})$ occurs at an asymmetric configuration. A similar analysis of bipartite entanglement between two quantum emitters placed inside a rolled-up ENZ waveguide has been previously analyzed[23]. It is interesting to note some important differences in the concurrence results between such an ENZ waveguide and the proposed MNZ metamaterial. First, the $(C_{ss})$ values for both cases are comparable when $r_{12} = 0.5\lambda_0$ (i.e., the emitters are within less than $\lambda_0$) with only slightly lower $(C_{ss})$ for the MNZ metamaterial. However, at longer distances such as $r_{12} = \lambda_0, 1.5\lambda_0, 5\lambda_0, 10\lambda_0, 15\lambda_0$ the achievable $(C_{ss})$ values for the MNZ metamaterial surpass those of the ENZ waveguide. Remarkably, the $(C_{ss})$ value achievable in the ENZ waveguide even at a short dipole-dipole separation of $r_{12} = 1.5\lambda_0$ falls short of the MNZ metamaterial with a large separation of $r_{12} = 15\lambda_0$ (i.e., once again one order of magnitude enhancement in the entanglement range). This further confirms that the proposed MNZ metamaterial is far superior to the ENZ waveguide in mediating an entanglement between two quantum emitters over a long range—much longer than the free-space wavelength.

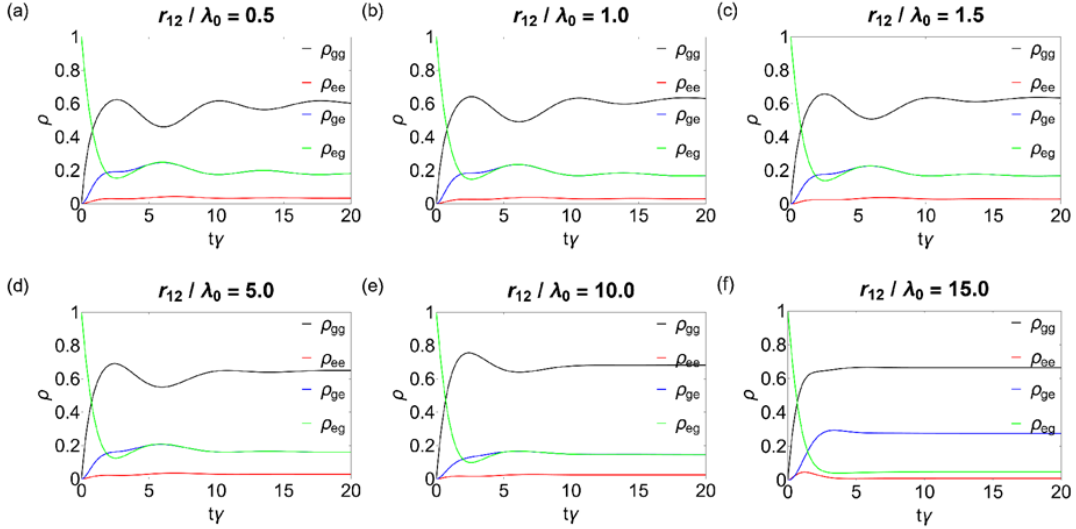

**Figure 7.** Time evolution of the diagonal elements of the density matrix at different dipole-dipole separations $r$ of **a)** $0.5\lambda_0$, **(b)** $\lambda_0$, **(c)** $1.5\lambda_0$, **(d)** $5\lambda_0$, **(e)** $10\lambda_0$, and **(f)** $15\lambda_0$ inside the MNZ metamaterial, evaluated at the maximum concurrence values for each separation.

Figure 7 displays the time evolution of the probabilities for the basis states (i.e., diagonal elements of the density matrix) using the same dipole-dipole separations as in Figure 6. They are obtained at the Rabi frequencies ($\Omega_1$ and $\Omega_2$) which give maximum concurrence for each dipole-dipole separation. Clearly, starting from the initial condition $\rho_{eg} = 1$, all the density matrix elements tend to reach the steady state at $\gamma t = 20$, consistent with the steady-state assumption in Fig. 6. Towards the steady state, we observe the following changes in the density matrix elements with increasing separations between the dipoles from $0.5\lambda_0$ (Fig. 7a) to $15\lambda_0$ (Fig. 7f) or equivalently with the decreasing $C_{ss}$. First, the probability $\rho_{gg}$ of finding both qubits at the ground state slightly increases, while the probability $\rho_{ee}$ of finding both in the excited state slightly decreases. Second, although the single-excitation probabilities $\rho_{ge}$ and $\rho_{eg}$ overlap in Figs. 7a-e, they are also slightly reduced with the decreased $C_{ss}$, and then dissociate in Fig. 7f, where the separation is $15\lambda_0$.

With the steady state density matrix of the present system, we can calculate the experimentally relevant photon-photon intensity correlations at zero time delay $g^{(2)}_{12}(0)$ as[39]

$$g^{(2)}_{12}(0) = \frac{\langle \sigma_1^\dagger \sigma_2^\dagger \sigma_2 \sigma_1 \rangle}{\langle \sigma_1^\dagger \sigma_1 \rangle \langle \sigma_2^\dagger \sigma_2 \rangle} = \frac{\rho_{ee}}{(\rho_{eg} + \rho_{ee})(\rho_{ge} + \rho_{ee})}. \tag{13}$$

For the case of a plasmonic ENZ waveguide[21], it has been demonstrated that the value of the zero time delay second order correlation function $g^{(2)}_{12}(0)$ for two-level systems can transition from bunching to antibunching $[g^{(2)}_{12}(0) = 0]$ depending on the value of the $\Omega_1$ and $\Omega_2$, and across varying interdipole separations. The antibunching signature has been associated with a high degree of entanglement[39]. In Fig. 8, using Eq. 13 we calculate the photon-photon correlations for the two two-level systems in the diamond MNZ metamaterial corresponding to Fig. 6. We observe that the regions of higher concurrence (shades of orange in Fig. 6) generally tend to correlate with the smaller values of $g_{12}^{(2)}(0)$ (shades of blue in Fig. 8). However, the concurrence for the present system better captures the degree of entanglement than $g_{12}^{(2)}(0)$. These findings are consistent with the literature[21]. We also note that with the increasing separation between the quantum emitters, the parameter space for antibunching shrinks as expected, consistent with Fig. 6. All materials for this section are available online[53].

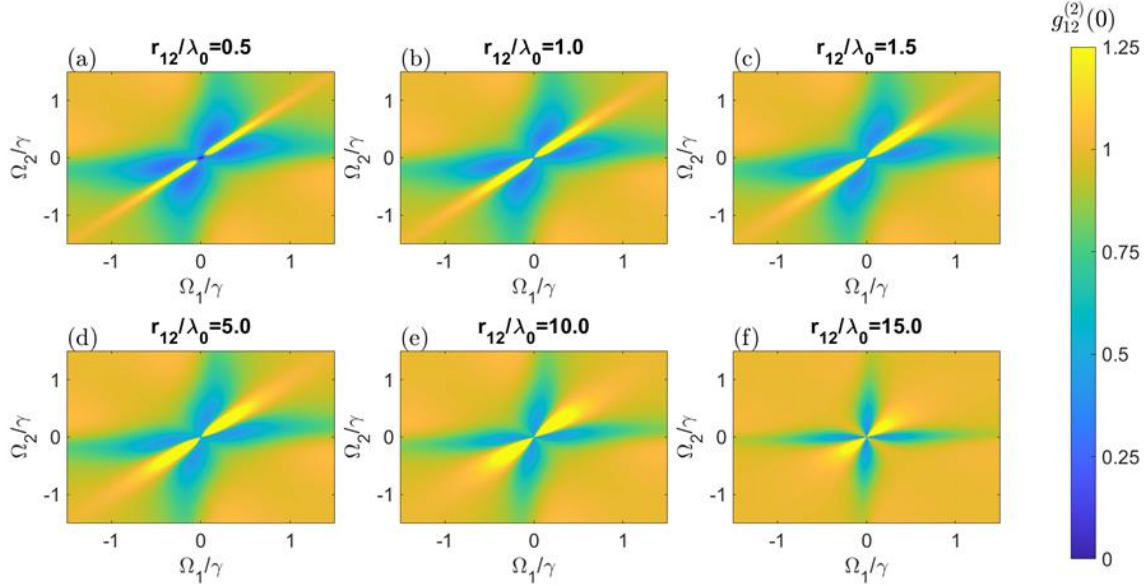

**Figure 8.** Steady state zero time delay second order correlation function $g_{12}^{(2)}(0)$ as a function of normalized Rabi frequencies $\Omega_1/\gamma$ and $\Omega_2/\gamma$ corresponding to the two dipoles in Fig. 6 separated by (**a**) $0.5\lambda_0$, (**b**) $\lambda_0$, (**c**) $1.5\lambda_0$, (**d**) $5\lambda_0$, (**e**) $10\lambda_0$, and (**f**) $15\lambda_0$.

## 5 Discussion about experimental implementation

The above discussion is purely theoretical, including both analytical and numerical developments. However, our work stands as close as possible to experiments. Therefore, we would like to briefly discuss the potential of this theoretical work for experimental realization by discussing fabrication, characterization, and potential imperfections in the system.

Many solid-state quantum emitters experience spectral inhomogeneity in their optical transitions. This is generally due to local strain variations in the diamond induced via fabrication[54]. A way to work around this effect is by using a Raman transition. A common technique in atomic, molecular, and optical physics is to induce a stimulated Raman transition between two meta-stable states, also known as a two-photon transition[55] . The spectra of these transitions contain two components — one broad peak corresponding to spontaneous emission at frequency ν and a narrow Raman peak corresponding to ν − Δ, where Δ is the frequency of the detuning. For relatively large values of Δ, we can use this narrow Raman peak to represent the transition we are interested in.

It is possible to pump the device out-of-plane with a Ti:Sapphire laser detuned both above (the "dressing" laser) and below (the "pump" laser) 737 nm to excite a Raman transition near the ZPL in the SiV and measure the emission in-plane via a tapered waveguide. To measure the cooperative decay rate Γ, one can record the temporal profiles of either forward or backward atomic emission in the MNZ mode.

One can scan a Ti:Sapphire laser over a 700 GHz range centered around 406.8 THz at steps of around 100 MHz. One can record fluorescence counts in the phonon sideband around the ZPL as a function of frequency to acquire all emitter resonances in the device. SiV centers can be ionized from the SiV state to either the SiV0 or SiV+2 charge states. If using the Raman transition in the far detuned scheme, the latter two states are "dark" and do not emit[56]. To correct for ionization, one can apply a 532 nm laser pulsed on the order of microseconds. This is called the "regenerating laser" in the setup. It returns the centers back to the SiV state.

Figure 9 shows a simplified schematic of this measurement. For the output waveguide, it is possible to either use an angle-etched diamond waveguide or a diamond waveguide suspended by etching from underneath the substrate. Additionally, MNZ and photonic bandgap designs are both possible with airhole configurations in diamond, and thus the entire structure.  It will likely be necessary to generate the short pump pulse (about 10 ns FWHM) to be able to fully decipher the enhanced decay rate.

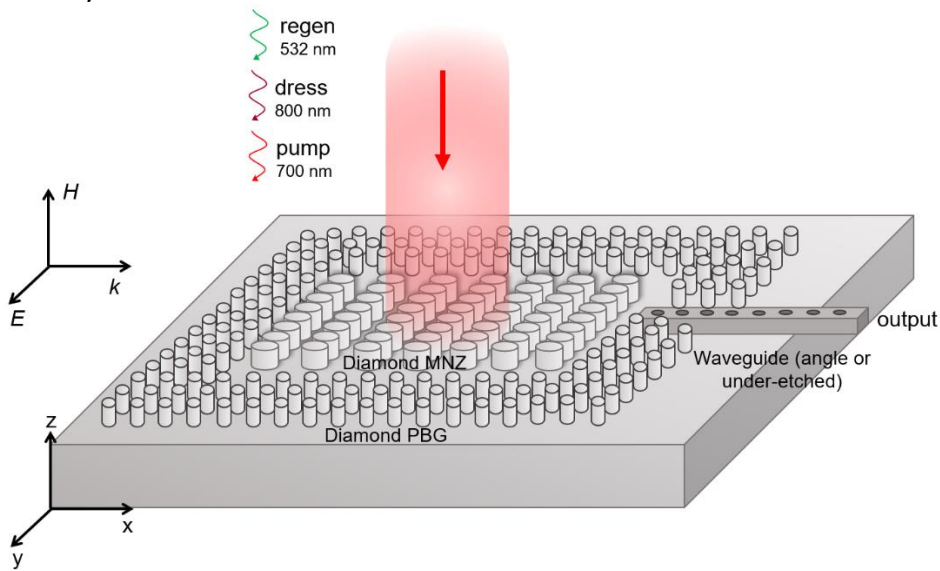

**Figure 9.** Schematic of MNZ material and PBG on the sides with pumping out of plane and measuring in-plane with a waveguide.

Traditional fabrication in single crystal bulk diamond is a challenging process that introduces specific requirements when both the nanostructures and the bulk are the same material.  For example, etching waveguides onto bulk diamond requires angle-etching of the sides of the waveguides to prevent additional leakage into the substrate[56].

Fortunately, recent research developments in diamond thin films have significantly improved the ease of fabrication and experimental feasibility of nanophotonic platforms with color vacancy centers in diamond. For example, recent developments in fabricating nanophotonic devices and photonic crystals from thin film diamond on silicon substrates show high-yield and use conventional planar fabrication techniques[57].

There are additional benefits to near-zero refractive index materials afforded to us due to the relaxed phase matching conditions which allow spatial uncertainties in the implantation of the vacancy centers in the diamond pillars. In the plot of the effective refractive index in Figure 2c, there is an approximately 20 nm region in which the effective refractive index of the MNZ material is at a minimum near zero, and throughout the entire simulation region of 100 nm the absolute effective refractive index is less than 0.5. As Maxwell's equations scale linearly with wavelength, small linear deviations of both pitch and radius will shift the wavelength of the effective constitutive parameters. While fabrication uncertainties (improper dosing of charge in electron beam lithography, under/over ion etching of devices), are not necessarily linear, small deviations in the pitch and radius of the photonic crystal within the range of near-zero refractive index behavior will not cause significant alterations in the effective refractive index.

To further demonstrate the relative robustness of the MNZ material design, we performed a simulation using COMSOL Multiphysics of our metamaterial with a pitch and radius of 505 nm and 115 nm, respectively, subjected to random perturbations of 2 nm and 5 nm on the cylinder radius (Figure 10). These simulations demonstrate that under these perturbations, we still maintain an effective refractive index close to zero around 737 nm wavelength.

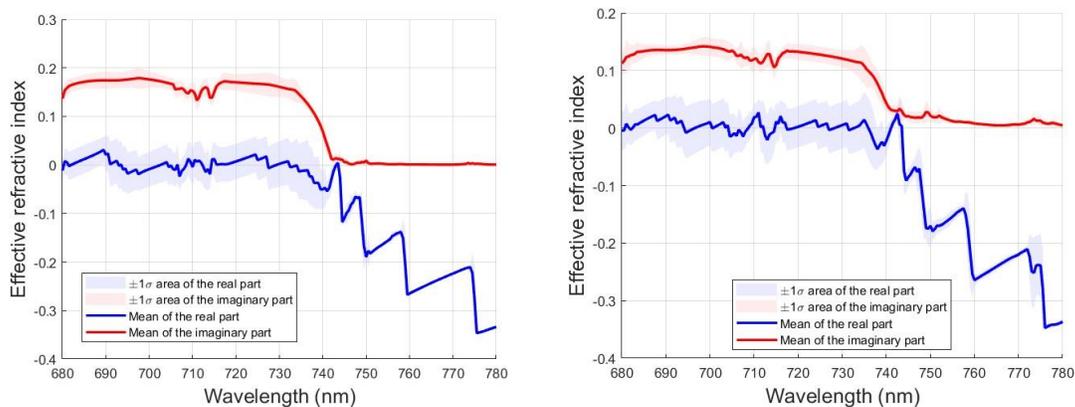

**Figure 10:** Effective refractive index of a 2D square lattice of diamond pillars ($n = 2.4064$) with a radius $r$ centered around 115 nm and randomly varied within ±2 nm (left) and ±5 nm (right). The crystal consists of 11 unit cells with a period $a = 505 nm$, demonstrating near-zero index behavior around 737 nm.

The proposed scheme could also be translated to other quantum emitters and technologies such as quantum dots, quantum wells, perovskite emitters, atoms, and cavity QED systems. Similar architectures have been previously proposed [31,32,58–63] and the quantum master equation can still describe those systems. Furthermore, a key feature of the NZRI structures, as well as general photonic crystals and metamaterials, is their scalability in wavelength due to the scalability of Maxwell's equations. Silicon vacancy centers in diamond here are chosen as candidate quantum emitters due to considerable experimental interest in them for the past decade, however the simulation and design process to generate NZRI photonic crystals and metamaterials is easily generalizable for different materials and wavelengths. There are several candidate emitters that could be immediately used in place of silicon vacancy centers in diamond. Some examples include silicon zero index metamaterials with quantum dots, or erbium doped silicon[59] which has narrow optical transitions between 1535 nm and 1538 nm.

## 6 Conclusions

We explored the dynamics of quantum entanglement within a MNZ metamaterial structure, emphasizing the temporal and spatial characteristics of concurrence. Our findings demonstrate that the MNZ structure significantly enhances entanglement between two quantum emitters beyond the capabilities of plasmonic ENZ waveguides and 2D all-dielectric ENZ platforms.

First, the concurrence maintained within the MNZ structure shows a robust enhancement, sustaining high levels of entanglement over distances up to 17 free-space wavelengths or approximately 12.5 µm. This is a substantial improvement – one order of magnitude – compared to both vacuum systems and ENZ waveguides.
More precisely, compared to free space, where cooperative decay enhancement significantly drops outside the range of a spatial wavelength to approximately the single emitter decay rate, with our MNZ material, we observe nearly a 16.5-fold spatial enhancement in the transient concurrence before we converge to the free space concurrence value. Extrapolating the results presented with ENZ plasmonic waveguide channels in [23] to regions extending further than one micron, we observe a 2.96-fold enhancement in both cooperative enhancement and steady-state concurrence. Compared to the rectangular ENZ waveguides in [21], we observe an 8.25-fold spatial enhancement in both the transient and steady-state concurrence as we increase the inter-emitter separation.
Moreover, through the application of an external pump source, we extended the transient concurrence into a steady state, allowing for prolonged entanglement. The analysis reveals that the maximum steady-state concurrence is achieved under antisymmetric pumping configurations, except at the largest separation distances, where asymmetric configurations prevail.
   Our comparative analysis further highlights that the MNZ metamaterial consistently surpasses the ENZ waveguide in terms of the maximum achievable concurrence across all studied separations. Even at shorter separations, where MNZ had only slightly lower concurrence, the longer-range performance proves far superior, reinforcing the MNZ structure's efficacy in mediating long-range entanglement.

Finally, the time evolution of the system's density matrix elements towards steady state supports our conclusions. The probabilities of the ground and excited states adjust in accordance with the increasing separation between dipoles, reflecting the decreasing concurrence. The value of the zero time delay second order correlation function $g_{12}^{(2)}(0)$ indicates antibunching signature correlated with the high degree of concurrence.

In summary, the MNZ metamaterial offers a highly effective medium for sustaining quantum entanglement over extended distances and time periods. These findings pave the way for future innovations in quantum communication and computation, leveraging the unique properties of MNZ structures to facilitate robust and long-range quantum entanglement.

Supplementary information accompanies the manuscript on the Light: Science & Applications website (http://www.nature.com/lsa).

*Acknowledgements*

The authors would like to thank Humeyra Caglayan for discussions about M. Issah thesis work. A.D is a Research Fellow of the Fonds de la Recherche Scientifique – FNRS. M.L. is a Research Associate of the Fonds de la Recherche Scientifique – FNRS. This research used resources of the "*Plateforme Technologique de Calcul Intensif (PTCI)*" (http://www.ptci.unamur.be) located at the University of Namur, Belgium, which is supported by the FNRS-FRFC, the Walloon Region, and the University of Namur (Conventions No. 2.5020.11, GEQ U.G006.15, 1610468, RW/GEQ2016 et U.G011.22). The PTCI is member of the "*Consortium des Équipements de Calcul Intensif (CÉCI)*" (http://www.ceci-hpc.be). This work was partially supported by the United States Army Research Office (ARO) under MURI grant (W911NF2420195).

*Author Contributions*

O.M. derived the analytical modeling and performed the numerical simulations under the guidance and supervision of L.V., M.L., and E.M. O.M., L.V., E.M., and M.L. performed the underlying analysis of the results. S.N. performed the steady state concurrence and time evolution under the supervision of D.G. A.D. performed simulations related to the photonic crystal modeling under the supervision of M.L. O.M., M.L., and D.G. wrote the manuscript with input from all the authors. All the authors subsequently took part in the revision process and approved the final copy of the manuscript.

*Conflict of Interest statement.*

The authors declare no competing interests.

***

# Supplementary information: Long-range quantum entanglement in dielectric mu-near-zero metamaterials

Olivia Mello[1], Larissa Vertchenko[2,], Seth Nelson[3], Adrien Debacq[4], Durdu Guney[1,5], Eric Mazur[1], Michaël Lobet[1,4,*]

[1]John A. Paulson School of Engineering and Applied Sciences, Harvard University, 9 Oxford Street, Cambridge, MA 02138, USA
[2] Sparrow Quantum, 2100 Copenhagen, Denmark
[3]Physics Department, Michigan Technological University, Houghton, Michigan, USA
[4] Department of Physics and Namur Institute of Structured Materials, University of Namur, Rue de Bruxelles 51, 5000 Namur, Belgium
[5] Electrical & Computer Engineering Department, Michigan Technological University, Houghton, Michigan, USA

*Corresponding author: michael.lobet@unamur.be

## 1. Imaginary parts of effective permittivity, permeability and effective index

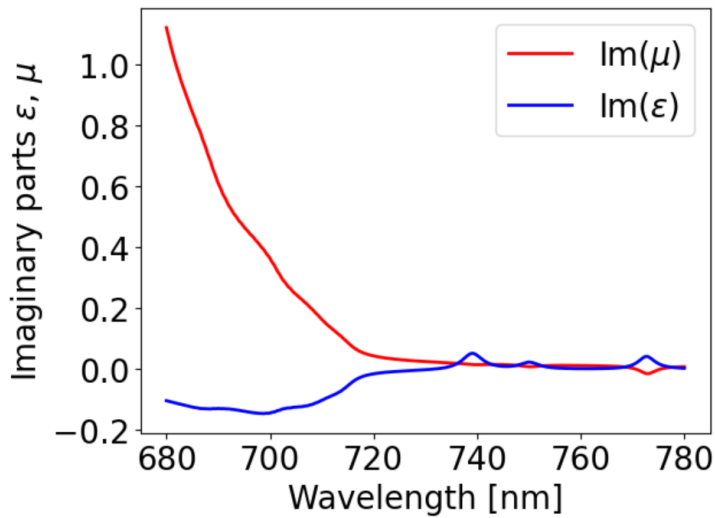

**Figure S1.** Imaginary parts of the effective permittivity and permeability corresponding to figure 2.

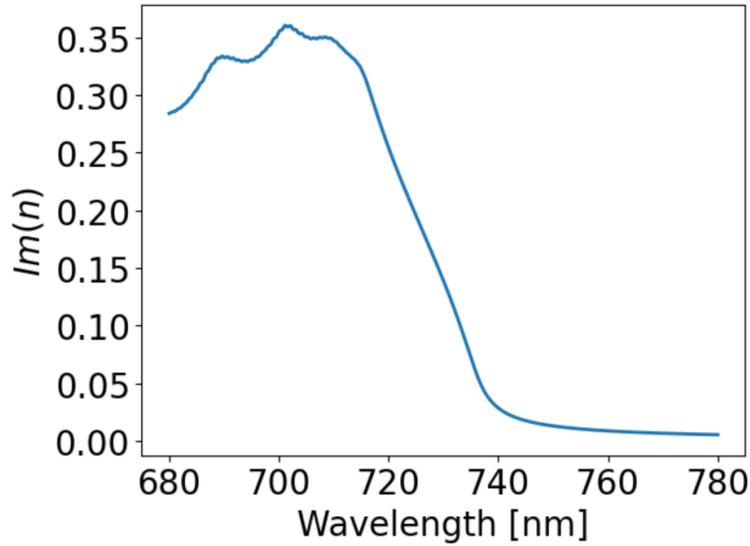

**Figure S2.** Imaginary parts of the effective refractive index corresponding to figure 2.

    To double-check the consistency of the retrieved parameters using the parameter retrieval methods, we calculated the effective index $n_{eff}$ calculated by averaging the phase advance between each pillar for 10 periods of the metamaterial using Ansys Lumerical FDTD. This agreement with $Re(\pm\sqrt{\varepsilon_r\mu_r})$ confirms the validity of the photonic crystal as a MNZ metamaterial structure.

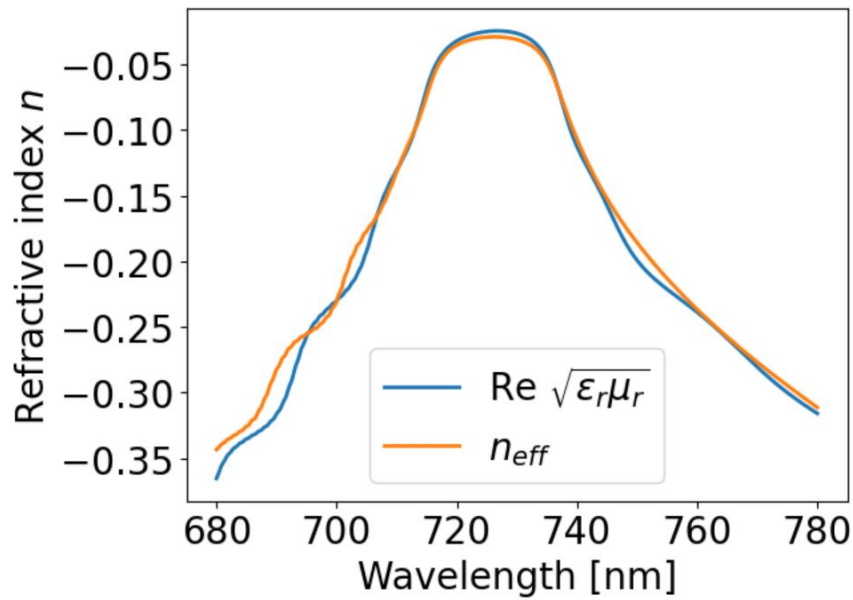

**Figure S3.** Comparison between the real part of the retrieved index $Re(\pm\sqrt{\varepsilon_r\mu_r})$ and the effective index $n_{eff}$ calculated using FDTD.

## 2. Master Equations for the Quantum Dynamics of the Emitters in the Presence of Pump Lasers

The time evolution of the density matrix $\rho(t)$ in the presence of potential term (Eq. 12 in the main text) is described by the following set of 16 equations using the standard basis of $|0\rangle = |g1, g2\rangle, |1\rangle = |g1, e2\rangle, |2\rangle = |e1, g2\rangle$ and $|3\rangle = |e1, e2\rangle$, and assuming a zero detuning $\Delta_i = 0$.

$$\frac{\partial \rho_{00}}{\partial t} = i\Omega_1(\rho_{20} - \rho_{02}) + i\Omega_2(\rho_{10} - \rho_{01}) + \gamma_{11}(\rho_{22} + \rho_{11}) + \gamma_{12}(\rho_{12} + \rho_{21})$$

$$\frac{\partial \rho_{01}}{\partial t} = i\Omega_1(\rho_{21} - \rho_{03}) + i\Omega_2(\rho_{11} - \rho_{00}) + \gamma_{11}\left(\rho_{23} - \frac{1}{2}\rho_{01}\right) + \gamma_{12}\left(\rho_{13} - \frac{1}{2}\rho_{02}\right) - ig_{12}\rho_{02}$$

$$\frac{\partial \rho_{02}}{\partial t} = i\Omega_1(\rho_{22} - \rho_{00}) + i\Omega_2(\rho_{12} - \rho_{03}) + \gamma_{11}\left(\rho_{13} - \frac{1}{2}\rho_{02}\right) + \gamma_{12}\left(\rho_{23} - \frac{1}{2}\rho_{01}\right) - ig_{12}\rho_{01}$$

$$\frac{\partial \rho_{03}}{\partial t} = i\Omega_1(\rho_{23} - \rho_{01}) + i\Omega_2(\rho_{13} - \rho_{02}) - \rho_{03}\gamma_{11}$$

$$\frac{\partial \rho_{10}}{\partial t} = i\Omega_1(\rho_{30} - \rho_{12}) + i\Omega_2(\rho_{00} - \rho_{11}) + \gamma_{11}\left(\rho_{32} - \frac{1}{2}\rho_{10}\right) + \gamma_{12}\left(\rho_{31} - \frac{1}{2}\rho_{20}\right) + ig_{12}\rho_{20}$$

$$\frac{\partial \rho_{11}}{\partial t} = i\Omega_1(\rho_{31} - \rho_{13}) - i\Omega_2(\rho_{10} - \rho_{01}) + \gamma_{11}(\rho_{33} - \rho_{11}) - \frac{1}{2}\gamma_{12}(\rho_{12} + \rho_{21}) + ig_{12}(\rho_{21} - \rho_{12})$$

$$\frac{\partial \rho_{12}}{\partial t} = i\Omega_1(\rho_{32} - \rho_{10}) + i\Omega_2(\rho_{02} - \rho_{13}) - \rho_{12}\gamma_{11} + \gamma_{12}\left(\rho_{33} - \frac{1}{2}(\rho_{11} + \rho_{22})\right) + ig_{12}(\rho_{22} - \rho_{11})$$

$$\frac{\partial \rho_{13}}{\partial t} = i\Omega_1(\rho_{33} - \rho_{11}) - i\Omega_2(\rho_{12} - \rho_{03}) - \frac{3}{2}\rho_{13}\gamma_{11} + \rho_{23}\left(ig_{12} - \frac{1}{2}\gamma_{12}\right)$$

$$\frac{\partial \rho_{20}}{\partial t} = i\Omega_1(\rho_{00} - \rho_{22}) + i\Omega_2(\rho_{30} - \rho_{21}) + \gamma_{11}\left(\rho_{31} - \frac{1}{2}\rho_{20}\right) + \gamma_{12}\left(\rho_{32} - \frac{1}{2}\rho_{10}\right) + ig_{12}\rho_{10}$$

$$\frac{\partial \rho_{21}}{\partial t} = i\Omega_1(\rho_{01} - \rho_{23}) + i\Omega_2(\rho_{31} - \rho_{20}) - \rho_{21}\gamma_{11} + \gamma_{12}\left(\rho_{33} - \frac{1}{2}(\rho_{11} + \rho_{22})\right) + ig_{12}(\rho_{11} - \rho_{22})$$

$$\frac{\partial \rho_{22}}{\partial t} = -i\Omega_1(\rho_{20} - \rho_{02}) + i\Omega_2(\rho_{32} - \rho_{23}) + \gamma_{11}(\rho_{33} - \rho_{22}) - \frac{1}{2}\gamma_{12}(\rho_{12} + \rho_{21}) + ig_{12}(\rho_{12} - \rho_{21})$$

$$\frac{\partial \rho_{23}}{\partial t} = -i\Omega_1(\rho_{21} - \rho_{03}) + i\Omega_2(\rho_{33} - \rho_{22}) - \frac{3}{2}\rho_{23}\gamma_{11} + \rho_{13}\left(ig_{12} - \frac{1}{2}\gamma_{12}\right)$$

$$\frac{\partial \rho_{30}}{\partial t} = -i\Omega_1(\rho_{32} - \rho_{10}) - i\Omega_2(\rho_{31} - \rho_{20}) - \rho_{30}\gamma_{11}$$

$$\frac{\partial \rho_{31}}{\partial t} = i\Omega_1(\rho_{11} - \rho_{33}) - i\Omega_2(\rho_{30} - \rho_{21}) - \frac{3}{2}\rho_{31}\gamma_{11} - \rho_{32}\left(\frac{1}{2}\gamma_{12} + ig_{12}\right)$$

$$\frac{\partial \rho_{32}}{\partial t} = -i\Omega_1(\rho_{30} - \rho_{12}) + i\Omega_2(\rho_{22} - \rho_{33}) - \frac{3}{2}\rho_{32}\gamma_{11} - \rho_{31}\left(\frac{1}{2}\gamma_{12} + ig_{12}\right)$$

$$\frac{\partial \rho_{33}}{\partial t} = -i\Omega_1(\rho_{31} - \rho_{13}) - i\Omega_2(\rho_{32} - \rho_{23}) - 2\rho_{33}\gamma_{11}$$

Here $\gamma_{11}$ is the single qubit decay rate and $\gamma_{12}$ is the cooperative decay rate. As the density matrix is Hermitian, the terms $\rho_{ij}(t)^* = \rho_{ji}(t)$. To solve for the steady state behavior of the system one can set the system $\partial \rho / \partial t = 0$ and solve for the eigenvalues of the matrix. Solving this secular equation gives us the steady-state density matrix elements $\rho_{SS}$ and the steady-state concurrence $C_{SS}$. Alternatively, the steady-state results can be approximated by letting the system evolve over a sufficiently long-time duration (e.g., $\gamma t = 90$ as in Fig. 6 in the main text, see also Fig. 7).

Refs. 8 and 23 of the main text use the quantum master equation and provide some limited numerical results for two emitters in plasmonic waveguides under one or two pumps leading to steady-states (e.g., concurrence as a function of time under a single pump with varying Rabi frequencies or two pumps with a few specific configurations of the Rabi frequencies, and steady-state concurrence for a short range with a few specific configurations of the Rabi frequencies). Although more detailed numerical analyses of the steady-state concurrence and zero-time delay second order correlation function were provided in Refs. 21 and 23 of the main text, they fall short of providing the general coupled differential equations above for the density matrix elements of the system in the presence of pumps to describe the full quantum dynamics of the two-qubit system under different initial conditions. Here we included those equations alongside a simplified analytical expression for the zero-time-delay second order correlation function in Eq. 13 in the main text as an entanglement measure relevant to experiments. Since the above equations are general in that they do not consider specific initial conditions, they can be directly employed to implement various quantum tasks.

### 3. Discussion about the advantage of MNZ vs ENZ vs EMNZ for coupling

The goal of this annex is to clarify the advantage of MNZ design compared to ENZ or EMNZ designs for coupling radiation into the system.

First, coupling is only possible along normal direction because of Snell's law of refraction. Indeed, if $n_2 = 0$ for the near-zero refractive index media to couple in, the only way to satisfy $n_1 sin\theta_1 = n_2 sin\theta_2$ from an incident medium different from a NZI ($n_1 \neq 0$) is to come at normal incidence ($\theta_1 = 0°$).

As we show below, it implies that the wave-vector must be zero for a plane wave propagating in a metamaterial with MNZ property.

From Maxwell's equations, this arises from the sourceless divergence of the electric field, which implies that the wavevector $\vec{k}$ must be orthogonal to the electric field $\vec{E}$:
$$\vec{\nabla} \cdot \vec{E} = 0 \implies \vec{k} \cdot \vec{E} = 0$$
Furthermore, since the permeability is zero in an MNZ medium, the electric field becomes irrotational, leading to:
$$\vec{\nabla} \times \vec{E} = 0 \implies \vec{k} \times \vec{E} = 0.$$
Thus, the only way the wavevector can be both parallel and orthogonal to the electric field is if the wavevector $\vec{k}$ is zero.

To couple with the zero-wavevector of a 2D photonic crystal, two coupling possibilities exist. First, in-plane coupling follows the wavevector defined at the Γ-point, aligning with the direction of periodicity (see Opt. Express 31, 26565-26576 (2023)[1]). Second, out-of-plane coupling involves a wave propagating perpendicular to the periodicity. If the periodicity is in the *xy*-plane, a wave propagating along the *z*-direction results in a zero wavevector in the *xy*-plane (see Nature 608, 692–698 (2022)[2]).

MNZ design better couples light in over ENZ design[3] because of the diverging impedance $Z = \sqrt{\frac{\mu}{\varepsilon}}$ in the ENZ case[4]. Nevertheless, in the case of a 2D MNZ photonic crystal, in-plane coupling is weak due to the impedance approaching zero, which leads to high reflectance at the zero refractive index frequency. However, above the bandgap ($R \to 1$) but still in the near zero refractive index regime (e.g. between 740-760 nm), light is transmitted according to Fabry-Perot oscillation, enabling in-plane coupling (Figure R2a). Further optimizations are required to maximize this coupling.

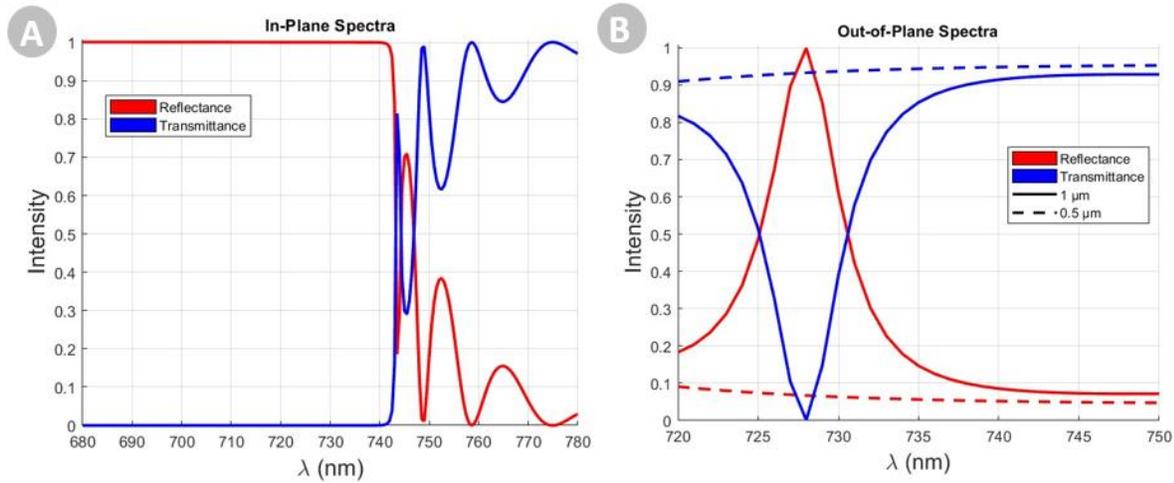

**Figure S4 :** (a) In-plane transmittance and reflectance spectrum of a 2D square lattice of diamond pillars ($n = 2.4064$, radius $r = 115$nm). The lattice consists of 11 unit cells with a period $a = 505\ nm$, showing MNZ behavior around a wavelength of 737 nm. (b) Out-of-plane transmittance and reflectance spectrum of a 2D square lattice of diamond pillars for different thicknesses. The lattice size is 5 unit cells with a period.

In contrast, out-of-plane coupling does not suffer from this Fabry-Perot oscillations since the MNZ behaviour is confined to the plane. The efficiency of out-of-plane coupling depends on the thickness of the crystal (Figure R2b) and can be optimized as described in the paper (see Light Sci Appl 10, 10 (2021)[5]).

In conclusion, coupling to the MNZ photonic crystal can occur both in-plane and out-of-plane.

Comparing to an EMNZ is a different problem. The impedance of an EMNZ structure responds according to a L'Hospital rule at the NZI limit and reaches $Z = \sqrt{\frac{\frac{d\mu}{d\omega}}{\frac{d\varepsilon}{d\omega}}}$ [4]. Therefore, a careful engineering of dispersion relation of both effective $\varepsilon$ and $\mu$ since there could have their independent slope variation as we reproduced the results from [6] here below (Figure S5):

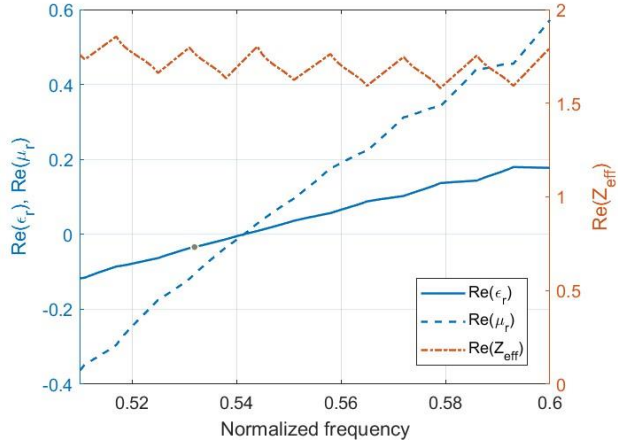

**Figure S5 :** Real permittivity and real permeability (left) reproduced from the structure simulated in [6], figure 2. It clearly shows that the slope of the permittivity and permeability can be different, leading to an effective impedance different from unity.

Nevertheless, EMNZ materials can be considered as a candidate for performance. The figure S6 below demonstrates the cooperative enhancement of a TM-polarized EMNZ material. We calculate this cooperative enhancement in the same manner as in Figure 5a of the manuscript, sweeping a line current dipole source across the high symmetry points of the EMNZ material while keeping another dipole source at the center of the material fixed. This metamaterial has a pitch of $a = 473\ nm$ and a pillar radius is $r = 113\ nm$. We calculate the cooperative enhancement by calculating the ratio of enhancement of the imaginary component of the electric field Ez for the two dipoles as they increase in separation to the field of the single dipole. This metamaterial offers a similar level of cooperative enhancement as the MNZ material in the manuscript.

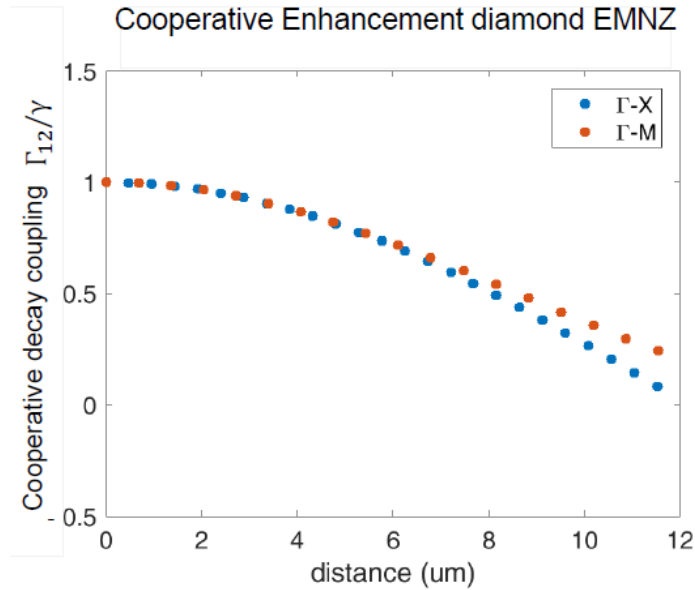

**Figure S6 :** The cooperative decay rate coupling $\Gamma_{12}/\gamma$ for a diamond EMNZ metamaterial in both the $\Gamma - X$ and $\Gamma - M$ directions.

The main advantage of using ENZ (TM polarization) or MNZ (TE polarization) is due to the Purcell enhancement one receives when the triply-degenerate Dirac cone at the Gamma point of EMNZ materials is detuned into the parabolic dispersion at the Gamma point that we observe in ENZ and MNZ materials. This parabolic dispersion and resulting band edge at the Gamma point introduces a significant decrease in the group velocity of the light at that point, which in turn increases the local density of states and thus the Purcell factor. The figure below shows the Purcell enhancement we observe in the MNZ material at 737 nm (see Figure 3a of the manuscript).

While this does not affect the overall cooperative enhancement, this does generate a greater overall emitted power and signal. The additional broadening of the linewidth occurring due to emission at the bandedge of either an MNZ or ENZ material also helps increase the indistinguishability of individual silicon vacancy centers emitting near the zero-phonon line.